\documentclass{JCIN22}
\usepackage{stfloats}

\begin{document}

\ArticleType{Research paper}
\Year{2022}

\title{LLM4AMC: Adapting Large Language Models for Adaptive Modulation and Coding}

\author{Xinyu Pan}
\author{Boxun Liu}
\author{Xiang Cheng}
\author{Chen Chen}

\maketitle

{\bf\textit{Abstract---}Adaptive modulation and coding (AMC) is a key technology in 5G new radio (NR), enabling dynamic link adaptation by balancing transmission efficiency and reliability based on channel conditions. However, traditional methods often suffer from performance degradation due to the aging issues of channel quality indicator (CQI). Recently, the emerging capabilities of large language models (LLMs) in contextual understanding and temporal modeling naturally align with the dynamic channel adaptation requirements of AMC technology. Leveraging pretrained LLMs, we propose a channel quality prediction method empowered by LLMs to optimize AMC, termed LLM4AMC. We freeze most parameters of the LLM and fine-tune it to fully utilize the knowledge acquired during pretraining while better adapting it to the AMC task. We design a network architecture composed of four modules, a preprocessing layer, an embedding layer, a backbone network, and an output layer, effectively capturing the time-varying characteristics of channel quality to achieve accurate predictions of future channel conditions. Simulation experiments demonstrate that our proposed method significantly improves link performance and exhibits potential for practical deployment.\\[-1.5mm]

}

\barefootnote{
This work was supported in part by the National Natural Science Foundation of China under Grant 62125101 and Grant 62341101; in part by the New Cornerstone Science Foundation through the Xplorer Prize.

X. Y. Pan, B. X. Liu, X. Cheng, C. Chen. State Key Laboratory of Photonics and Communications, School of Electronics,
Peking University, Beijing 100871, China (e-mail: 2501213461@stu.pku.edu.cn; boxunliu@stu.pku.edu.cn; xiangcheng@pku.edu.cn; c.chen@pku.edu.cn).\\\indent
}


\section{Introduction}

With the advent of 5G, dynamic link adaptation, multimodal information fusion, advanced modulation and coding schemes, and other key enabling technologies have become essential for meeting the demanding requirements of low latency and high data rate communications\textsuperscript{\cite{ref1,ref2}}. As a fundamental component of link adaptation, adaptive modulation and coding (AMC) dynamically selects an appropriate modulation and coding scheme (MCS) based on the channel quality indicator (CQI)\textsuperscript{\cite{ref3}}. The CQI is typically derived from the signal-to-interference-plus-noise ratio (SINR) observed on the channel, while the MCS is composed of two elements, the modulation order and the coding rate. In essence, AMC establishes a mapping from channel quality to the optimal MCS, aiming to strike a balance between link reliability and transmission efficiency. Specifically, a higher modulation order and coding rate can lead to increased data throughput, but also result in a higher block error rate (BLER). When the channel condition is favorable, characterized by a high SINR, the system can adopt a more aggressive MCS with higher modulation and coding rate. Conversely, under poor channel conditions, a more conservative MCS is necessary to satisfy the BLER constraint and maintain link robustness. 

Currently, solutions to the AMC problem can be broadly categorized into three types. The first consists of traditional rule-based approaches, which are widely implemented in practical communication systems. The second includes deep learning-based methods that leverage data-driven models to capture complex mappings. The third category comprises reinforcement learning (RL) -based approaches, which focus on optimizing decision strategies through interactive learning and have emerged as a significant research direction in AMC.

Traditional AMC techniques typically employ a dual-loop control mechanism that combines inner loop link adaptation (ILLA) and outer loop link adaptation (OLLA). ILLA, also known as the fixed lookup table approach, establishes a strict one-to-one mapping between each CQI and a corresponding SINR range, which is then directly mapped to a specific MCS\textsuperscript{\cite{ref4}}. In contrast, OLLA introduces a closed-loop control mechanism by adjusting the estimated SINR based on ACK/NACK feedback, thereby refining the link adaptation process. This technique was first proposed in Ref. \cite{ref5} and further analyzed in depth in Ref. \cite{ref6}. Building upon conventional AMC methods, Ref. \cite{ref7} provides a detailed description of the AMC implementation procedure in 5G systems. The study generates SINR–BLER curves via Gaussian channel simulations and optimizes the correction factor within the effective SINR mapping algorithm, thereby improving the accuracy of the effective SINR estimation module. Additionally, Ref. \cite{ref8} proposes a CQI-free AMC algorithm that relies solely on ACK/NACK feedback to dynamically adapt the MCS. However, traditional AMC methods, due to their simple mapping relationships and limited input information (typically only SINR), often struggle to adapt the MCS selection strategy dynamically based on real-time channel conditions, resulting in performance degradation.

In recent years, the rapid development of deep learning technologies has provided new insights for optimizing AMC strategies. By leveraging their powerful nonlinear modeling capabilities and the ability to expand the input feature space (beyond relying solely on SINR), deep learning-based AMC approaches have effectively overcome the inherent limitations of traditional rule-based methods. Ref. \cite{ref9} proposes an AMC algorithm based on online deep learning (ODL), wherein neural network parameters are updated in real time to predict the success rate of each MCS under current channel conditions. The optimal MCS is then selected by jointly considering these predicted success rates and the spectral efficiency of each scheme. In Ref. \cite{ref10}, a feedback-free adaptive MCS selection framework is introduced, combining convolutional neural networks (CNNs) with long short-term memory (LSTM) networks. This method utilizes uplink channel state information (CSI) in multi-user scenarios to determine the appropriate downlink MCS, eliminating the need for explicit feedback.

Moreover, RL-based AMC has emerged as a significant research direction. Within the RL framework, the AMC problem is typically formulated as a Markov decision process (MDP)\textsuperscript{\cite{ref11}}, an approach adopted in early studies such as Refs. \cite{ref12} and \cite{ref13}. Subsequent research has extended RL-based AMC to more complex application scenarios. For instance, Ref. \cite{ref14} investigates AMC techniques in emergency power communication systems. Ref. \cite{ref15} proposes a joint optimization framework integrating AMC and resource allocation, and Ref. \cite{ref16} explores the coordinated scheduling of AMC with spatial multiplexing strategies. In addition, to address the challenge of CQI staleness, studies such as Refs. \cite{ref17, ref18} formulate the state space by integrating historical CQI or SINR measurements, prior MCS selections, and ACK/NACK feedback. The RL agent then selects the optimal action, i.e., the most appropriate MCS. Ref. \cite{ref19} reformulates the AMC task as a CQI prediction problem and applies a multi-armed bandit (MAB) model to estimate the current channel CQI, thereby improving the MCS selection strategy.

Although many existing studies have taken latency issues into account, most of them focus on optimizing the overall architecture of AMC. However, inspired by advances in channel prediction\textsuperscript{\cite{ref20}}, it is also feasible to incorporate channel quality prediction to improve the accuracy of CQI feedback, thereby enhancing the effectiveness of AMC decisions. In fact, the CQI prediction method proposed in Ref. \cite{ref19} can be regarded as a specific form of channel quality prediction. However, since each CQI value corresponds to a range of SINR levels rather than a precise measurement, predicting CQI inherently provides less accuracy improvement compared to directly predicting the SINR itself. 

Recently, large language models (LLMs) represented by GPT-4, Llama, and DeepSeek have demonstrated significant breakthroughs across various domains. They have achieved remarkable success not only in traditional AI fields such as natural language processing (NLP) and computer vision (CV) but have also been adapted to specialized areas like finance, healthcare, and education. By pretraining on massive datasets, these models acquire strong general modeling and generalization capabilities, opening up new possibilities for the optimization of communication systems. 
For instance, Ref. \cite{ref21} proposed a large language model-based channel prediction model, LLM4CP, which performs CSI prediction by fine-tuning a pretrained GPT-2 model. Similarly, Ref. \cite{ref22} also focuses on the channel prediction problem, aligning the network architecture, data processing methods, and optimization objectives with the pretraining process of large language models. This design enables the model to handle variable-length inputs and perform continuous multi-step predictions. Furthermore, Refs. \cite{ref23, ref24} apply fine-tuned large language models to millimeter-wave beam prediction and fluid antenna port prediction, respectively, both achieving higher prediction accuracy than traditional approaches. Ref. \cite{ref25} targets physical layer tasks and proposes LLM4WM, which addresses multiple CSI-related tasks. In addition, Ref. \cite{ref26} investigates LLM-driven CSI feedback for massive MIMO systems, demonstrating that pre-trained models can effectively compress and reconstruct CSI. Ref. \cite{ref27} explores the integration of LLMs with reconfigurable intelligent metasurface antenna systems, highlighting the potential of large language models to dynamically configure metasurfaces and enhance wireless communication performance.

Unlike existing approaches, this paper explores the application of LLMs to the task of channel quality prediction in single-input single-output (SISO) - orthogonal frequency division multiplexing (OFDM) systems, thereby effectively enabling AMC. Specifically, we design a channel quality prediction network based on a pretrained LLM, and fine-tune it to predict future SINR based on historical SINR measurements. The predicted SINR is then mapped to CQI values, addressing the CQI aging issue. To bridge the gap between traditional NLP tasks and channel quality prediction, we propose a network architecture composed of four modules, including a preprocessing layer, an embedding layer, a backbone network, and an output layer. During training, most of the model's parameters are frozen, and only a small portion is fine-tuned. Simulation results demonstrate the effectiveness of the proposed method and its potential for practical deployment. The main contributions of this paper are summarized as follows.

\begin{itemize}
    \item We propose an LLM-empowered channel quality prediction and AMC optimization method, LLM4AMC, to address the CQI aging issue in existing AMC schemes. 
    \item A dedicated network architecture is designed based on the characteristics of channel quality data. To the best of our knowledge, this is the first attempt to apply LLMs to the AMC task.
    \item Simulation results demonstrate that the proposed method can effectively enhance link performance. It also shows excellent performance in few-shot learning and generalization experiments, indicating strong potential for practical deployment.
\end{itemize}

\section{System Model}

This paper investigates a time division duplex (TDD) - single user (SU) - SISO - OFDM system. The proposed system model is inherently scalable and can be naturally extended to multi-user and multiple-input multiple-output (MIMO) scenarios.

\subsection{Traditional AMC Procedure}

\begin{figure}[htbp]
    \centering
    \includegraphics[width=0.45\textwidth, clip, trim=0 200 0 20]{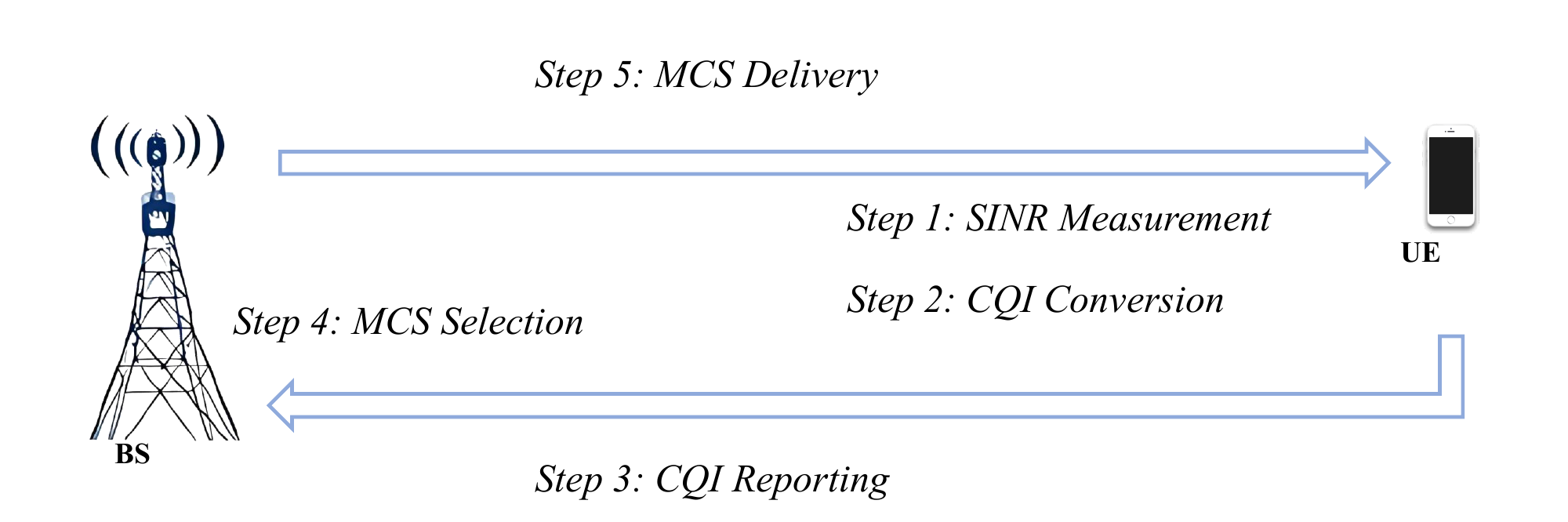}
    \caption{Illustration of the downlink AMC procedure}
    \label{fig:Illustration}
\end{figure}

Fig. ~\ref{fig:Illustration} illustrates the overall workflow of traditional AMC system. Without loss of generality, we focus on the downlink AMC problem. The user equipment (UE) first measures the SINR of the downlink signal and maps it to CQI for feedback to the base station (BS). The BS then selects an appropriate modulation and coding scheme based on the received CQI and transmits it to the user for data demodulation. It is worth noting that since this work focuses on channel quality prediction, the proposed method serves as an upper-layer enhancement module that can be applied to various AMC algorithms including both ILLA and OLLA. The performance gains are consistent across both schemes. Therefore, we select the ILLA method, which does not rely on ACK/NACK feedback, as a representative case for modeling and experimentation.

The user measures the channel quality at fixed intervals. Assuming an ideal measurement process, it is considered that accurate subcarrier SINR information can be obtained, denoted as $\bm{S}(t)  = \{S_1(t), S_2(t), \ldots, S_K(t)\}$, where K represents the total number of subcarriers.

In the 5G NR standard, to reduce feedback overhead, the user does not convert and feed back the SINR of each individual subcarrier to the BS as CQI. Instead, the SINR values are first combined into an effective SINR, denoted as $S_{eff}(t)$, which is then mapped to a CQI feedback. This process is referred to as effective SINR mapping (ESM). The general expression for ESM is given by

\begin{equation}
S_{eff}(t) = \beta f_{comp}^{-1}\left( \frac{1}{K} \sum_{i=1}^{K} f_{comp}\left( \frac{S_i(t)}{\beta} \right) \right)
\end{equation}

\noindent where $f_{comp}(\cdot)$ represents the ESM compression function, where different ESM algorithms employ different compression functions, and $f_{comp}^{-1}(\cdot)$ denotes its inverse function. The parameter $\beta$ serves as a tuning factor. The commonly used ESM algorithms include exponential effective SINR mapping (EESM) and mutual information effective SINR mapping (MIESM). In this study, we select EESM for effective SINR mapping, whose compression function is an exponential function defined as

\begin{equation}
f_{comp}(x) = \exp(-x)
\end{equation}

After obtaining the effective SINR, the UE converts the effective SINR to CQI through a standardized mapping function as follows.

\begin{equation}
CQI(t) = f_{CQI}(S_{eff}(t))
\end{equation}

\noindent where $f_{CQI}(\cdot)$ is established based on pre-simulated SINR-BLER curves, where each discrete CQI index corresponds to a specific SINR range. In this study, the mapping relationship is derived from the 5G system simulation model data in Ref. \cite{ref28}. Ultimately, the UE feeds back the CQI to the BS, where each CQI index corresponds to a unique MCS that determines both the modulation order and the code rate.

The MCS selected by the BS at time $t$ is denoted as $MCS(t) = \{Q(t), R(t)\}$, where $Q(t) \in \mathcal{M}$ represents the selected modulation order at time $t$, $\mathcal{M} = \{2:QPSK, 4:16QAM, 6:64QAM, 8:256QAM\}$ defines the set of supported modulation orders and their corresponding modulation schemes, $R(t) \in (0,1]$ denotes the selected code rate, with the open interval explicitly excluding zero-rate transmission scenarios.

The downlink SINR at time $t$ is denoted as $\bm{S}(t)$, and the corresponding BLER of the link can be expressed as

\begin{equation}
BLER(t) = f_{MCS}(MCS(t), \bm{S}(t))
\end{equation}

\noindent where $f_{MCS}(\cdot)$ represents the MCS-dependent BLER characteristic function obtained through link-level simulations.

Therefore, the achievable link throughput $T$ is given by

\begin{equation}
T(t) = (1 - BLER(t)) \cdot Q(t) \cdot R(t) \cdot TBS
\end{equation}

\noindent where TBS denotes the Transport Block Size.

\subsection{MCS Selection Based on Channel Quality Prediction}

The ILLA algorithm adopted in the above system model follows a fundamental principle, namely selecting the MCS that maximizes throughput while satisfying the BLER constraint. In theory, the MCS chosen by ILLA should be optimal under ideal conditions. However, the performance of AMC in real-world systems often falls short of this theoretical expectation. The primary reason for this discrepancy lies in the mismatch between the reported CQI and the actual channel conditions. Typically, inaccuracies in CQI can be attributed to three main factors, namely measurement error, hardware imperfection and feedback delay. Among these, measurement and hardware errors are intrinsic to the system and are generally not the focal point of AMC optimization. In contrast, feedback delay exerts a more profound impact on link performance, particularly in high-mobility scenarios where rapid channel variation exacerbates the mismatch between the reported CQI and the true channel state.

From the perspective of the AMC procedure, the BS needs to transmit reference signals to the UE for channel quality measurement, which introduces forward transmission delay. Subsequently, the UE completes the channel quality measurement and feeds back the CQI to the BS, inevitably incurring reverse transmission delay. This implies that the CQI information used for MCS selection corresponds to a channel state that lags behind the actual channel conditions at the time when the MCS is applied by at least one full round-trip feedback delay. Furthermore, when considering processing delays at both the UE and BS, as well as queuing delays, this latency can extend to several transmission time intervals (TTIs). In high-mobility scenarios, such delay leads to severe CQI staleness issues, resulting in a mismatch between the selected MCS and the current channel quality, which degrades link performance. Therefore, if the channel quality at the MCS application time can be predicted based on historical channel information and mapped back to a CQI for feedback to the BS, AMC performance can be significantly improved.

Therefore, we introduce a channel quality prediction module to improve the existing AMC technique. Specifically, a prediction step is added after Step 1, where the future SINR is estimated based on historical SINR values. The predicted SINR is then converted into a CQI and fed back to the BS. The subsequent procedure remains the same as in traditional AMC.

\section{LLM for Adaptive Modulation and Coding}

In the system described previously, the CQI feedback incurs a non-negligible latency denoted as $T_{\tau}$. To address the aging issue of CQI, we introduce a channel quality prediction module. When the user obtains the measurement results $\bm{S}(nT_s)$ at the n-th measurement cycle, the model is required to predict the SINR at time $nT_s + T_{\tau}$ based on historical SINR data. This can be formally expressed as

\begin{align}
\bm{S}(nT_s + T_{\tau}) 
&= f_{SINR}\big( 
  \bm{S}(nT_{\tau}),\ \bm{S}((n-1)T_{\tau}),\ \ldots, \notag \\
&\quad\quad\ \bm{S}((n-L+1)T_{\tau}) \,\big|\, \theta \big)
\end{align}

\noindent where $\theta$ denotes the model parameters.

\begin{figure*}[t]  
    \centering
    \includegraphics[width=0.9\textwidth, clip, trim = 50 150 50 150]{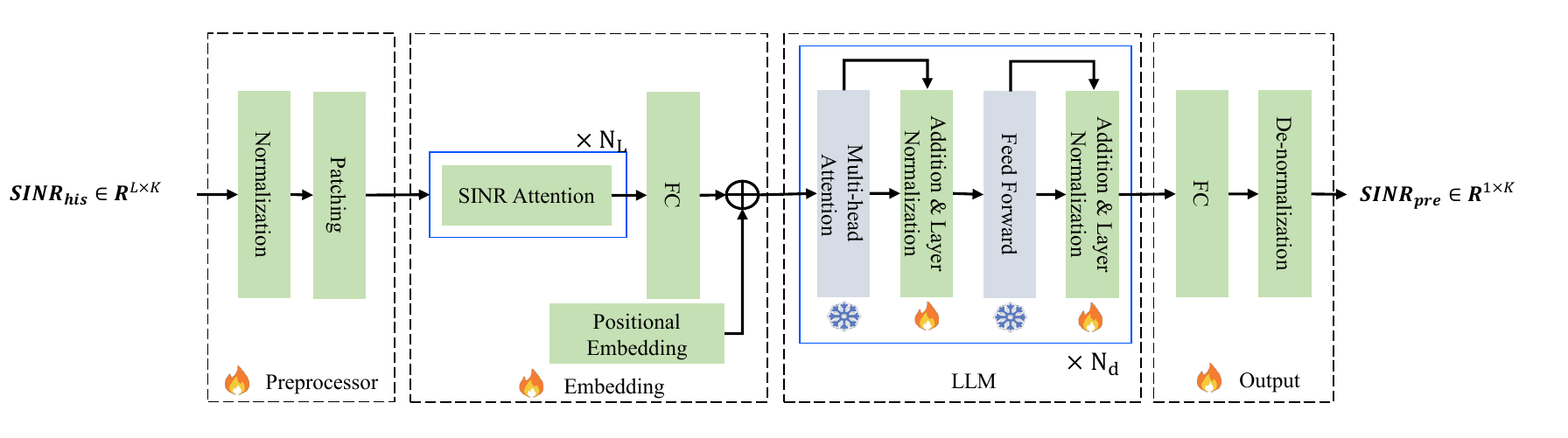}
    \caption{Architecture of the channel quality prediction network based on fine-tuned large language models}
    \label{fig:net_arch}
\end{figure*}

To address this challenge, we propose the network architecture illustrated in Fig. ~\ref{fig:net_arch}. The framework comprises four key modules, including a preprocessing layer, an embedding layer, a backbone network, and an output layer. In the diagram, trainable modules are annotated with red flame icons, while frozen parameter blocks are marked with blue snowflake icons. The implementation details of each module will be elaborated in subsequent sections.

\subsection{Preprocessing Layer}

In the preprocessing layer, the input SINR matrix $\bm{S} \in \mathbb{R}^{L \times K}$ is first normalized as follows.

\begin{equation}
\hat{\bm{S}} = \frac{\bm{S} - \mu}{\sigma}
\end{equation}

\noindent where $\mu$ denotes the mean of the input data and $\sigma$ represents its variance.

Subsequently, to effectively capture the temporal characteristics of channel quality while reducing computational complexity, the network performs non-overlapping block processing on the data\textsuperscript{\cite{ref29}}. Specifically, the continuous data is partitioned into fixed-length segments of size $N$, where the total number of blocks can be expressed as

\begin{equation}
L' = \left\lceil \frac{L}{N} \right\rceil
\end{equation}

\noindent Zero-padding is applied to the trailing elements to obtain the output $\bm{S}_P \in \mathbb{R}^{L' \times N \times K}$.

\subsection{Embedding Layer}

\begin{figure}[htbp]
    \centering
    \includegraphics[width=0.45\textwidth, clip, trim =200 50 300 70]{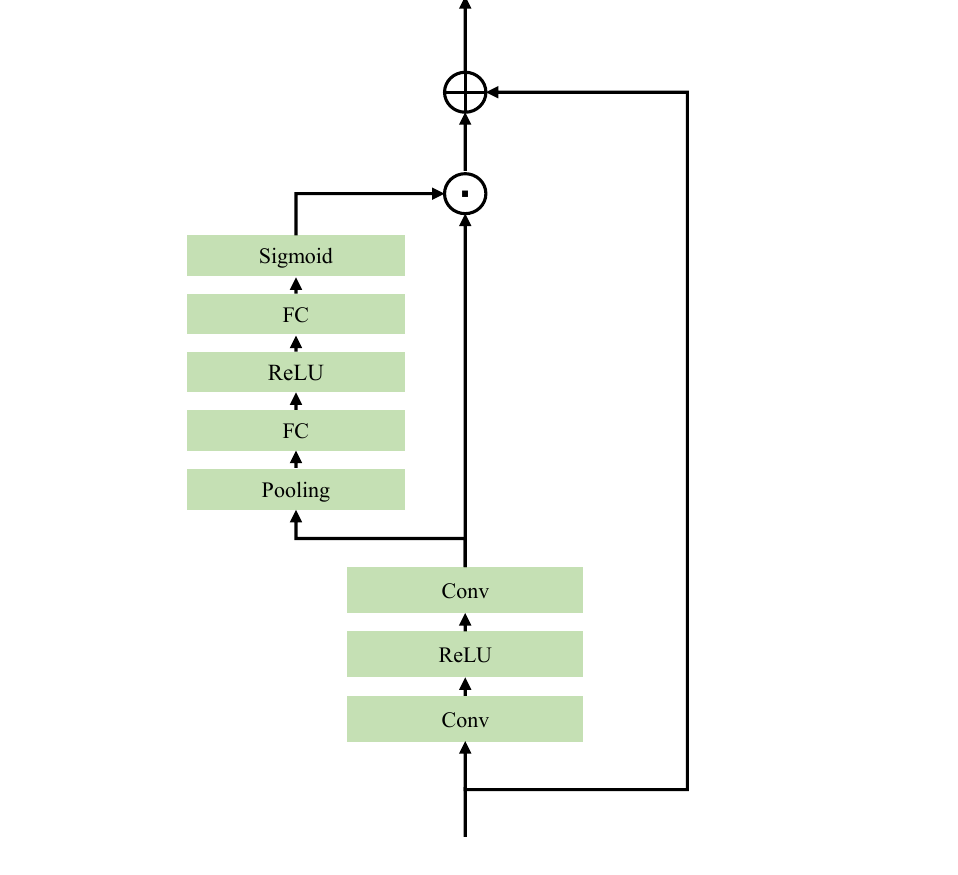}
    \caption{Architecture of the proposed SINR attention module.}
    \label{fig:embedding}
\end{figure}

The embedding layer is designed to align high-dimensional SINR features with textual representations through three sequential operations, namely feature extraction, attention enhancement, and dimensional transformation. Notably, the feature extraction and attention enhancement stages collectively constitute the SINR Attention Module, whose architecture is illustrated in Fig. ~\ref{fig:embedding}.

First, the feature extraction stage initially processes the segmented SINR values through a dual-layer convolutional architecture, with the mathematical representation given by Eq. (9).

\begin{equation}
\bm{S}_f = Conv(ReLU(Conv(\bm{S}_P))) \in \mathbb{R}^{L' \times N \times K}
\end{equation}

\noindent where $Conv(\cdot)$ denotes the 2D convolution operation and $ReLU(\cdot)$ represents the ReLU activation function. Through these two consecutive convolutional operations, the model effectively extracts temporal features of channel quality.

Subsequently, a squeeze-and-excitation (SE) channel attention mechanism\textsuperscript{\cite{ref30}} is introduced during the attention enhancement stage, formulated as

\begin{equation}
\bm{S}_{SE} = SE(\bm{S}_f) \in \mathbb{R}^{L' \times 1 \times 1}
\end{equation}

\noindent where $SE(\cdot)$ denotes the SE module. Specifically, the SE module first applies global average pooling to $\bm{S}_f$ to obtain $\bm{S}_{GAP}$, expressed as

\begin{equation}
\bm{S}_{GAP}[i] = \frac{1}{NK} \sum_{j=1}^{N} \sum_{k=1}^{K} \bm{S}_f[i,j,k]
\end{equation}

Subsequently, channel-wise dependencies are learned through fully-connected (FC) layers with a bottleneck structure, where $r$ denotes the reduction ratio and the hidden layer dimension is set to $L'/r$. A ReLU activation function is applied between the two fully-connected layers. Finally, the attention weights $\bm{S}_{SE} \in \mathbb{R}^{L' \times 1 \times 1}$ are generated via a sigmoid activation function. The scaled features $\bm{S}_{Sca} \in \mathbb{R}^{L' \times N \times K}$ are then obtained by element-wise multiplication of the weights with the input features, formulated as

\begin{equation}
\bm{S}_{Sca}[i,:,:] = \bm{S}_{SE}[i] \times \bm{S}_f[i,:,:]
\end{equation}

Finally, the output is obtained through residual connection.

\begin{equation}
\bm{S}_{output} = \bm{S}_{Sca} + \bm{S}_P
\end{equation}

This attention mechanism can be iteratively applied to fully extract features. Therefore, the output of the SINR attention module can be expressed as

\begin{equation}
\bm{S}_{SA} = SA^{(N_{SA})}\left(\bm{S}_P\right)
\end{equation}

\noindent where $SA(\cdot)$ denotes the SINR attention process described in Eqs. (9)-(14), and $SA^{(N_{SA})}(\cdot)$ represents the cyclic application of the SINR self-attention mechanism for $N_{SA}$ iterations.

After completing SINR feature extraction, to align with the input dimensionality requirements of the LLM, $S_{SA}$ undergoes dimension transformation. First, $\bm{S}_{SA}$ is reshaped from a 3D tensor to a 2D matrix, yielding $\bar{\bm{S}}_{SA} \in \mathbb{R}^{L' \times NK}$. Subsequently, a fully-connected layer projects $\bar{\bm{S}}_{SA} \in \mathbb{R}^{L' \times NK}$ to $\tilde{\bm{S}}_{SA} \in \mathbb{R}^{L' \times d_{model}}$, where $d_{model}$ denotes the hidden layer dimension of the pretrained LLM.

To compensate for the LLM's innate lack of sequential perception (as transformer architectures are permutation-invariant), we introduce positional encoding. Ref. \cite{ref31} demonstrates that trainable positional embeddings yield marginal improvements. Consequently, we implement the $sin-cos$ encoding scheme

\begin{equation}
\left\{
\begin{aligned}
\bm{S}_{PE}(2i,j) &= \sin\left( \frac{j}{10000^{\frac{2i}{d_{model}}}} \right) \\
\bm{S}_{PE}(2i+1,j) &= \cos\left( \frac{j}{10000^{\frac{2i}{d_{model}}}} \right)
\end{aligned}
\right.
\end{equation}

Thus, the output of the embedding layer is expressed as

\begin{equation}
\bm{S}_{EB} = \tilde{\bm{S}}_{SA} + \bm{S}_{PE}
\end{equation}

\subsection{Backbone Network}

In recent years, pretrained LLMs have demonstrated remarkable cross-domain generalization capabilities. Although LLMs were originally designed for NLP tasks, recent studies\textsuperscript{\cite{ref21, ref22, ref23, ref24, ref25, ref26, ref27}} have shown that with appropriate fine-tuning, these models can be effectively extended to non-textual domains such as time-series forecasting.

In this work, through the preceding preprocessing and embedding layers, the model effectively aligns channel quality information with textual representations. Consequently, the raw channel quality data is transformed into abstract features, enabling full utilization of the LLM's powerful modeling capabilities. Subsequently, the embedded features $S_{EB}$ are fed into the LLM backbone network for deep feature extraction, formulated as

\begin{equation}
\bm{S}_{LLM} = LLM\left(\bm{S}_{EB}\right) \in \mathbb{R}^{L' \times d_{model}}
\end{equation}

\noindent where $LLM(\cdot)$ denotes the backbone network of the large language model.

For the LLM selection, we compared models from the GPT\textsuperscript{\cite{ref32}}, DeepSeek\textsuperscript{\cite{ref33}}, Llama\textsuperscript{\cite{ref34}}, and Qwen\textsuperscript{\cite{ref35}} series. Balancing prediction accuracy and inference latency, we ultimately adopted the Qwen2.5-0.5B model as our backbone network. The specific experimental procedures and conclusions are detailed in the following sections. During fine-tuning, to achieve an optimal trade-off between predictive performance and computational efficiency, we froze the weights of both the multi-head self-attention mechanisms and feed-forward neural networks, updating only the layer normalization parameters. This approach significantly reduced the number of trainable parameters.

\subsection{Output Layer}

The primary function of the output layer is to map the backbone network's output to the final prediction results. First, a two-layer fully connected network is employed for dimension transformation, expressed as

\begin{equation}
\bm{S}_{pre} = FC\left(FC\left(\bm{S}_{LLM}\right)\right) \in \mathbb{R}^{1 \times K}
\end{equation}

\noindent where $FC(\cdot)$ denotes the fully-connected layer. After obtaining the prediction results, $\bm{S}_{pre}$ undergoes denormalization to restore the original physical units, yielding the final predicted output

\begin{equation}
\hat{S} = \sigma \bm{S}_{pre} + \mu
\end{equation}

\subsection{Training Configuration}

To align with practical application scenarios, the model must maintain stable prediction performance across varying user velocities. Therefore, the training dataset must incorporate channel quality samples at multiple velocity points. During training, the ground-truth SINR values are known and denoted as $\bm{S}_{truth} \in \mathbb{R}^{1 \times K}$, and we employ normalized mean square error (NMSE) as the loss function, defined as

\begin{equation}
LOSS(\hat{\bm{S}}, \bm{S}_{truth}) = NMSE(\hat{\bm{S}}, \bm{S}_{truth})
\end{equation}

\section{Experiments}

This section demonstrates through simulations that the fine-tuned LLM-based channel quality prediction method effectively addresses the aging issues of CQI, thereby enhancing AMC technology.

\subsection{Simulation Parameter Configuration}

\subsubsection{Dataset Configuration}
This study employs the QuaDRiGa channel simulation platform\textsuperscript{\cite{ref36}} to generate 3GPP-compliant\textsuperscript{\cite{ref37}} CSI data. The system operates at a center frequency of 2.4 GHz ($f_c=2.4$ GHz) with a subcarrier spacing of 15 kHz ($\Delta f=15$ kHz), allocating 48 subcarriers per user ($K=48$), corresponding to a total bandwidth of 720 kHz.

For channel modeling, this study adopts the 3GPP-defined urban macrocell (UMa) non-line-of-sight (NLOS) propagation model. The BS's transmit power is set to 40 dBm ($P_{BS}=40$ dBm), with a noise power of -84 dBm ($N=-84$ dBm). For time-domain parameters, the TTI is 0.5 ms, the historical SINR window length is 16 ($L=16$), the SINR measurement period is $T_m=2$ TTI, and the CQI feedback delay is $T_d=2$ TTI.

The user mobility pattern follows a linear trajectory, with initial positions randomly distributed within the BS's coverage area. The training and validation sets consist of 72,000 and 8,000 samples, respectively, with user speeds uniformly distributed between 40-100 km/h. The test set comprises seven discrete speed points (40 km/h, 50 km/h, ..., 100 km/h), each containing 1,000 independent samples. Detailed channel simulation parameters are listed in Tab. ~\ref{tab:Channel_para}.

\begin{table}[htbp]
\centering
\scriptsize
\caption{Channel simulation parameters}
\label{tab:Channel_para}
\begin{tabular}{cc}
\toprule
\textbf{Parameter} & \textbf{Configuration} \\
\midrule
Scenario & UMa NLOS \\
Center Frequency ($f_c$) & 2.4 GHz \\
Subcarrier Spacing ($\Delta f$) & 15 kHz \\
Number of Subcarriers ($K$) & 48 \\
BS Transmit Power ($P_{BS}$) & 40 dBm \\
Noise Power ($N$) & $-84$ dBm \\
TTI & 0.5 ms \\
SINR History Window Length ($L$) & 16 \\
SINR Measurement Period ($T_m$) & 2 TTI \\
CQI Feedback Delay ($T_d$) & 2 TTI \\
\bottomrule
\end{tabular}
\end{table}

\subsubsection{Benchmark Models}
To thoroughly validate the effectiveness of the proposed model, this study selects several representative baseline models for comparison. First, the no prediction (NP) model directly uses the last measured SINR as the current prediction, quantifying the impact of CQI feedback delay on system performance. Second, three typical recurrent neural networks are selected for comparison, including standard RNN\textsuperscript{\cite{ref38}}, LSTM\textsuperscript{\cite{ref39}}, and GRU\textsuperscript{\cite{ref40}}, all configured with 4 layers while maintaining consistent loss functions and training procedures with the fine-tuned large language model.

\subsubsection{Network Architecture and Training Parameters}
The network configurations in our simulations are set as follows. In the preprocessing layer, the patch size is set to 4, denoted as $N=4$. The SINR attention module undergoes 4 iterations denoted as $N_{SA}=4$. The backbone network employs 6 layers of Qwen2.5 denoted as $N_{LLM}=6$. The primary network architecture and training parameters are detailed in Tab. ~\ref{tab:arch}.

\subsubsection{Performance Metrics}
The evaluation in this study comprises two categories, namely prediction accuracy metrics and AMC performance metrics. The former validates the model's predictive capability, while the latter demonstrates the performance gains brought by the proposed method to AMC technology. The NMSE is adopted to quantify the discrepancy between predicted and measured SINR values. For AMC performance evaluation, both BLER and throughput are employed, where BLER references simulation data from Ref. \cite{ref28}, and throughput is calculated using Eq. (5). The CQI table used in all simulation experiments is Table 5.2.2.1-2 from Ref. \cite{ref41}.

\begin{table}[htbp]
\centering
\scriptsize
\caption{Network architecture and training parameters}
\label{tab:arch}
\begin{tabular}{cc}
\toprule
\textbf{Parameter} & \textbf{Configuration} \\
\midrule
Patch Size ($N$) & 4 \\
SA Iterations ($N_{SA}$) & 4 \\
LLM Layers ($N_{LLM}$) & 6 \\
Batch Size & 512 \\
Training Epochs & 500 \\
Optimizer & Adam (betas = (0.9, 0.999)) \\
Initial Learning Rate & 0.001 \\
\bottomrule
\end{tabular}
\end{table}

\subsection{Simulation Results Analysis}

\subsubsection{Prediction Performance Across Different User Velocities} 

\begin{figure}[htbp]
    \centering
    \includegraphics[width=0.4\textwidth]{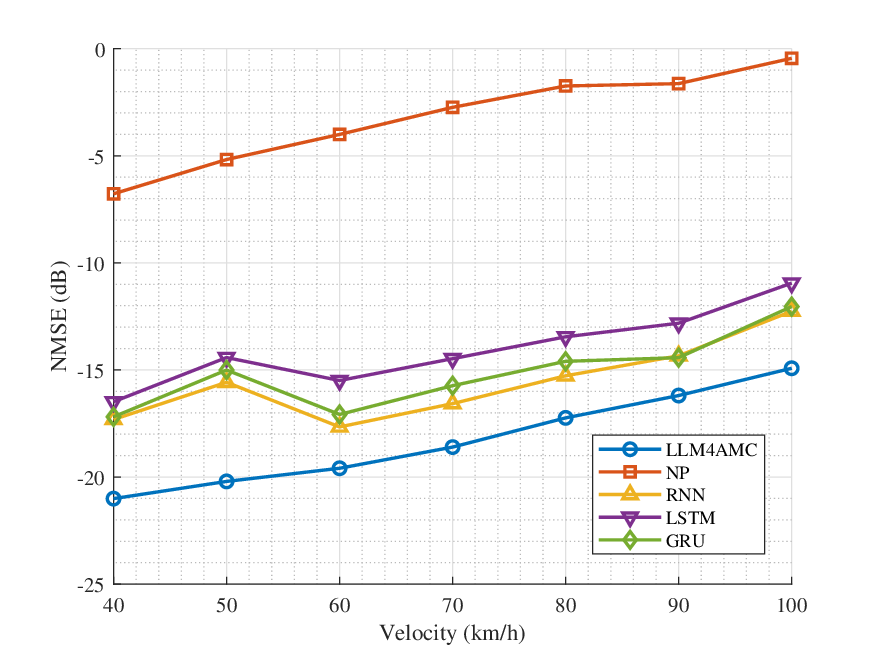}
    \caption{Comparison of prediction NMSE across different user velocities.}
    \label{fig:all_speed}
\end{figure}

Fig. ~\ref{fig:all_speed} compares the NMSE performance of various prediction models across user velocities (40-100 km/h), while Tab. ~\ref{tab:all_speed} presents the average BLER and throughput statistics for each model. The results demonstrate that the NP model exhibits significantly higher NMSE than other approaches, with the performance gap widening as velocity increases. In link-level simulations, the NP model achieves a BLER of 0.3097, substantially exceeding the target BLER of 0.1. These findings underscore the necessity of channel quality prediction for mitigating CQI aging issues.

\begin{table}[htbp]
\centering
\scriptsize
\caption{Comparison of average BLER and throughput across all speeds}
\label{tab:all_speed}
\begin{tabular}{ccccc c}
\toprule
\textbf{Metric} & \textbf{LLM4AMC} & \textbf{NP} & \textbf{RNN} & \textbf{LSTM} & \textbf{GRU} \\
\midrule
BLER & \textbf{0.1178} & 0.3097 & 0.1739 & 0.1865 & 0.1876 \\
Throughput (Mbps) & \textbf{8.0533} & 5.8331 & 7.8228 & 7.6475 & 7.7388 \\
\bottomrule
\end{tabular}
\end{table}

Moreover, compared to RNN, LSTM, and GRU models, LLM4AMC demonstrates significant performance advantages. The proposed method achieves the lowest NMSE across all velocity points, particularly under high-speed mobility (100 km/h) scenarios, where it improves prediction accuracy by 45.8\% over the suboptimal model. In terms of AMC performance metrics, the fine-tuned LLM approach also delivers optimal results. Compared to the NP model, it reduces BLER by 61.96\% and increases throughput by 38.06\%. When benchmarked against the suboptimal RNN model, it achieves a 2.95\% higher throughput and 32.26\% lower BLER. Notably, the simulations did not account for hybrid automatic repeat request (HARQ) mechanisms. In practical systems equipped with HARQ, the substantial BLER improvement would significantly reduce retransmissions, thereby yielding even more considerable throughput gains.

Furthermore, in order to evaluate the robustness of the proposed framework under extreme conditions, we consider the ultra-high-mobility scenario envisioned in 6G, where user speeds can reach up to 500 km/h. To this end, we construct a dataset consisting of 80,000 samples uniformly distributed within the velocity range of 300--500 km/h. The model is trained and compared against the NP baseline under the same setting. The results show that LLM4AMC achieves a NMSE of -14.86 dB, whereas the NP model reaches -12.74 dB, clearly demonstrating the superior prediction capability of our method in ultra-high-mobility scenarios. We also attempted to train lightweight models such as RNN, LSTM, and GRU, but they failed to converge. This indicates that due to the inherent complexity of ultra-high-speed environments, traditional small models are no longer applicable, while the proposed LLM4AMC provides a viable and effective solution.

The effectiveness of LLM4AMC can be attributed to two key factors. First, through large-scale pretraining, the LLM acquires strong temporal modeling capabilities, enabling it to accurately capture long-term dependencies in channel quality variations, which are often overlooked by conventional recurrent models. Second, compared with traditional RNN-based approaches, LLM4AMC incorporates a task-specific architectural design that explicitly aligns the input-output characteristics with the requirements of channel quality prediction. This design ensures that the pretrained knowledge is effectively transferred and adapted to the AMC task, resulting in superior performance across diverse and challenging mobility conditions.

\subsubsection{Noise Robustness Testing}

In practical systems, channel quality prediction inevitably suffers from SINR measurement errors, necessitating robust noise immunity in prediction models. During the training phase, all models undergo data augmentation with additive white Gaussian noise (AWGN) uniformly distributed between 15-25 dB. For testing, model robustness is evaluated by injecting noise of varying intensities into historical SINR data.

\begin{figure}[htbp]
    \centering
    \includegraphics[width=0.4\textwidth]{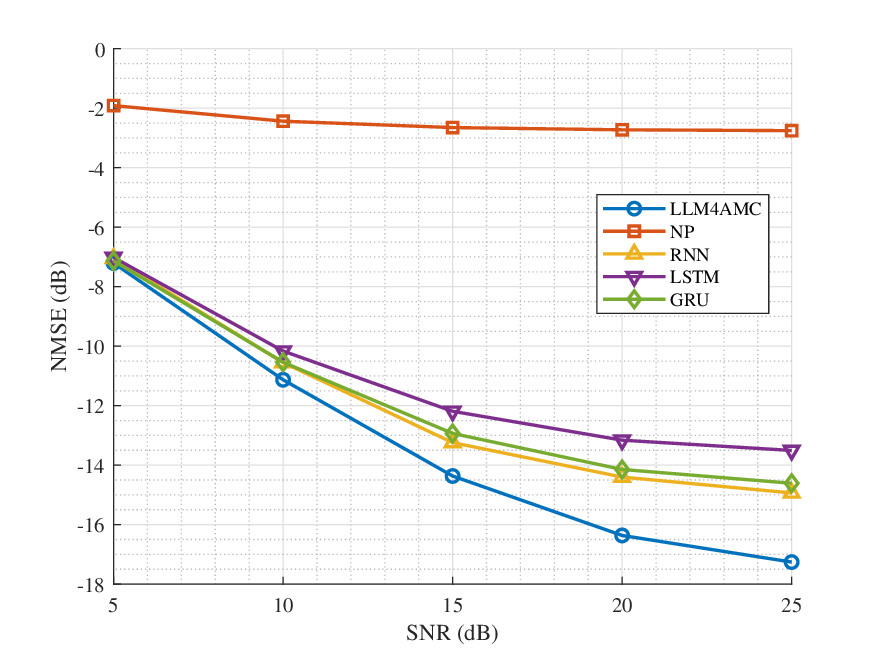}
    \caption{Prediction accuracy of different models under varying SNR conditions.}
    \label{fig:snr}
\end{figure}

As shown in Fig. ~\ref{fig:snr}, the experimental results demonstrate two key findings. First, the NMSE of all models decreases with increasing data SNR, eliminating the possibility that models learned noise patterns instead of SINR variation patterns. Second, the proposed method maintains lower NMSE across all SNR levels, highlighting its robustness advantage. These results not only confirm the reliability of our approach under measurement errors but also demonstrate its practical deployability in real-world systems.

\subsubsection{Few-Shot Learning Performance}

\begin{figure}[htbp]
    \centering
    \includegraphics[width=0.4\textwidth]{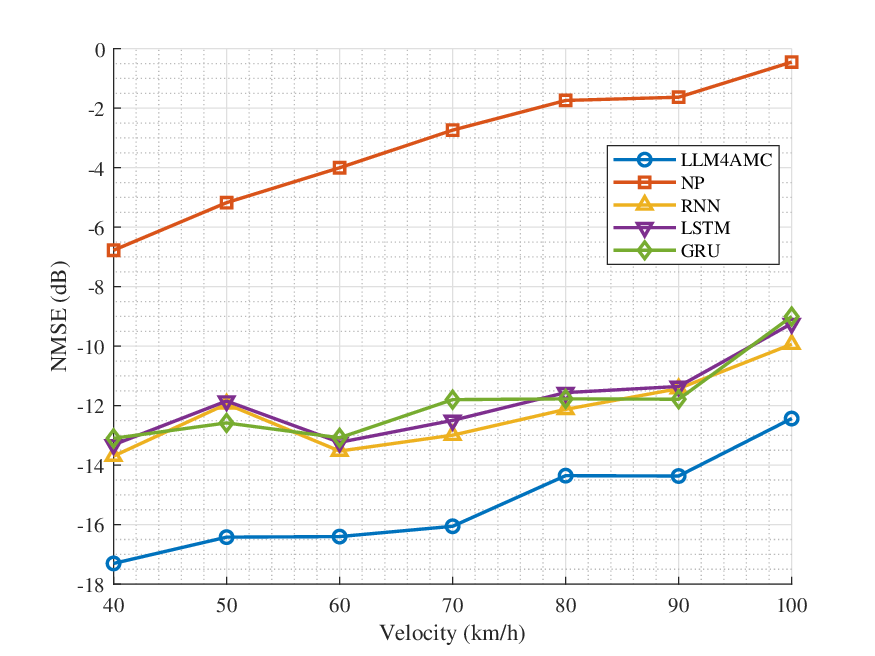}
    \caption{Few-shot learning performance.}
    \label{fig:few_shot}
\end{figure}

This study further evaluates the few-shot learning capabilities of different models. Specifically, each model is trained using only 10\% of the complete dataset, followed by performance evaluation on the test set. The experimental results are shown in Fig. ~\ref{fig:few_shot}.

The results demonstrate that under the extreme condition of using only 10\% of training samples, the fine-tuned LLM-based channel quality prediction model, while exhibiting performance degradation compared to full-data training, maintains significant advantages over baseline models. First, its prediction accuracy remains markedly superior to the NP model, proving the method's validity even in few-shot scenarios. Second, the fine-tuned LLM approach consistently achieves the lowest NMSE among all benchmark models. These findings confirm that although pretrained LLMs typically require massive parameters and substantial data, the proposed fine-tuning strategy, which updates only a minimal subset of parameters , effectively reduces data dependency while preserving practical applicability.

\subsubsection{Cross-Scenario Generalization Capability}
 This study further evaluates the cross-scenario generalization capability of each model. Specifically, both the proposed model and baseline models are trained on the UMa-NLOS scenario and subsequently tested on the Urban microcell (UMi)-NLOS scenario to assess their generalization performance. The NMSE, BLER, and throughput of each model are presented in Tab. ~\ref{tab:gen}.

\begin{table}[htbp]
\centering
\scriptsize
\caption{Generalization performance}
\label{tab:gen}
\begin{tabular}{cccccc}
\toprule
\textbf{Metric} & \textbf{LLM4AMC} & \textbf{NP} & \textbf{RNN} & \textbf{LSTM} & \textbf{GRU} \\
\midrule
NMSE (dB) & \textbf{-17.59} & -2.73 & -15.36 & -13.87 & -15.12 \\
BLER & \textbf{0.4747} & 0.6135 & 0.4959 & 0.5105 & 0.5148 \\
Throughput (Mbps) & \textbf{3.0905} & 2.0775 & 3.0412 & 2.9523 & 2.9751 \\
\bottomrule
\end{tabular}
\end{table}

The experimental results demonstrate that LLM4AMC exhibits strong cross-scenario generalization capability. Compared to the NP model, the proposed method maintains significant advantages in the UMi-NLOS scenario, with 22.63\% lower BLER and 48.76\% higher throughput, confirming its effectiveness. Furthermore, the fine-tuned LLM approach also outperforms other baseline models in cross-scenario generalization performance.

\subsubsection{Comparison of Different LLMs}
The network architecture we proposed is highly versatile and can flexibly accommodate any LLM as its backbone. However, in practical applications, differences in model size and pretraining data across various LLMs lead to significant variations in inference latency and adaptability to the AMC task. These differences substantially affect the overall performance and deployability of our method. Therefore, we compared several open-source LLMs, including Qwen, GPT, DeepSeek, and Llama, to identify the most suitable model for this task. In our experiments, we selected 8,000 training samples from the original dataset with a fixed user speed of 50 km/h, while keeping all other conditions unchanged. We then evaluated the system performance on an additional set of 1,000 new samples.

Considering the processing latency requirements of the model, the selected LLMs must not have excessively long inference times, and therefore their parameter sizes should also be moderate. After a comprehensive comparison, we selected four models for evaluation: Qwen2.5-0.5B, Qwen2.5-0.5B-Instruct, DeepSeek-R1-Distill-Qwen-1.5B, and Llama-3.2-1B. For brevity, these models are denoted in Tab. ~\ref{Tab.diff_llms} as Qwen, Qwen-Inst, DeepSeek, and Llama, respectively. To ensure fairness, the number of layers for all models was set to 6, i.e., $N_{\text{LLM}} = 6$. The parameter size, inference latency, and prediction NMSE of each model when used as the backbone are summarized in Tab. ~\ref{Tab.diff_llms}.

As shown, Qwen2.5-0.5B achieves the best prediction performance while also exhibiting the shortest inference time when used as the backbone, making it the most reasonable choice for the overall network architecture. Additionally, it is worth noting that we also compared Qwen2.5-0.5B with its instruction-tuned variant, Qwen2.5-0.5B-Instruct. The experimental results indicate that instruction tuning did not provide any performance gain for our network, and thus the instruction-tuned model was not selected.

\begin{table}[htbp]
\centering
\caption{Comparison of different LLMs used as backbone in terms of parameter size, inference time, and prediction performance}
\label{Tab.diff_llms}
\scriptsize 
\begin{tabular}{ccccc}
\toprule
Models & Parameters (M) & Inference Time (ms) & NMSE (dB) \\
\midrule
\textbf{Qwen} & 226.11 & 1.54 & \textbf{-22.29} \\
\textbf{Qwen-Inst} & 226.11 & 1.55 & -22.26 \\
\textbf{DeepSeek} & 748.16 & 5.74 & -21.84 \\
\textbf{Llama} & 628.32 & 4.81 & -21.27 \\
\bottomrule
\end{tabular}
\end{table}

\subsubsection{Ablation Experiments}
To comprehensively assess the effectiveness and design choices of our approach, we conducted three types of ablation studies, each serving a different purpose. The first focuses on model components, where we remove key modules to evaluate their individual contributions. The second targets the training data scale, in which the model is trained with different proportions of the dataset to investigate its data efficiency. The third examines fine-tuning strategies, comparing alternative adaptation methods to analyze their impact on performance. Together, these ablation experiments provide a holistic understanding of how architectural design, data availability, and fine-tuning choices influence the proposed framework.

For ablation on model components, we individually removed the SINR attention module, the patching module, and the LLM module from the network. The average NMSE, BLER, and throughput across all velocities are reported in Tab. ~\ref{Tab.module_ablation} Evidently, each of these modules is essential for achieving high performance.

\begin{table}[htbp]
\centering
\caption{Ablation study of LLM4AMC in terms of NMSE, BLER, and Throughput across all velocities}
\label{Tab.module_ablation}
\scriptsize
\begin{tabular}{ccccc}
\toprule
\textbf{Metrics} & \textbf{LLM4AMC} & \textbf{w/o SA} & \textbf{w/o Patching} & \textbf{w/o LLM} \\
\midrule
NMSE (dB) & \textbf{-17.75} & -16.31 & -17.69 & -17.28 \\
BLER     & \textbf{0.1178}  & 0.1613  & 0.1206  & 0.1247  \\
Throughput (Mbps) & \textbf{8.0533}  & 7.8314  & 7.9822  & 7.9736  \\
\bottomrule
\end{tabular}
\end{table}

For ablation on training data scale, we trained the model with different proportions of the dataset, including 20\%, 40\%, 60\%, 80\%, and the full dataset. The evaluation was performed on the same test set to ensure fairness. As shown in Tab. ~\ref{Tab.data_ablation}, the results demonstrate that the performance steadily improves with larger training data, highlighting the data efficiency of our framework. In addition to the NMSE values, we also report the NMSE increase, which measures how much the error increases compared with the best result obtained using the full dataset.

\begin{table}[htbp]
\centering
\caption{Ablation study of LLM4AMC with different training data scales}
\label{Tab.data_ablation}
\scriptsize
\begin{tabular}{cccccc}
\toprule
\textbf{Metric} & 20\% & 40\% & 60\% & 80\% & Full \\
\midrule
NMSE (dB) & -16.17 & -16.34 & -16.58 & -16.94 & \textbf{-17.75} \\
NMSE Increase  & 43.96\% & 38.3\% & 30.92\% &  20.65\%& \textbf{0.00\%}\\
\bottomrule
\end{tabular}
\end{table}

For ablation on fine-tuning strategies, we first consider fine-tuning only the LayerNorm layers, which is the method adopted in this work. We then examine three alternatives including updating all parameters through full fine-tuning, freezing all parameters without any adaptation, and fine-tuning both the LayerNorm and the MLP layers. The results are presented in Tab. ~\ref{Tab.ft_ablation}. Similar to the data-scale ablation, we also report the NMSE increase, where the comparison is made against the best result achieved by the LN-only strategy. This highlights that tuning only the LayerNorm layers preserves the pretrained knowledge most effectively while enabling efficient adaptation.

\begin{table}[htbp]
\centering
\caption{Ablation study of LLM4AMC with different fine-tuning strategies}
\label{Tab.ft_ablation}
\scriptsize
\begin{tabular}{ccccc}
\toprule
\textbf{Metirc} & LN & All Params & Frozen & LN + MLP \\
\midrule
NMSE (dB) & \textbf{-17.75} & -16.63 & -16.98 & -16.49 \\
NMSE Increase  & \textbf{0.00\%} & 29.29\% & 19.54\% &  33.7\% \\
\bottomrule
\end{tabular}
\end{table}

As shown in Tab.~\ref{Tab.ft_ablation}, the best performance is achieved by fine-tuning only the LayerNorm layers, which is the method adopted in this work. This strategy reaches an NMSE of -17.75 dB and serves as the baseline for comparison. When all parameters are updated through full fine-tuning, the performance clearly degrades despite the much larger number of trainable parameters. This suggests that excessive parameter updates may overwrite or disturb the valuable knowledge already embedded in the pretrained backbone, leading to suboptimal adaptation. Freezing all parameters also yields inferior results since the model cannot adjust to the downstream AMC task. Fine-tuning both the LayerNorm and the MLP layers shows moderate improvement over full fine-tuning and frozen settings, but still falls behind the LN-only strategy. These observations indicate that the pretrained LLM already encodes strong representations for communication tasks, and lightweight adaptation through LayerNorm alone is sufficient to align feature distributions to the target scenario while retaining the rich knowledge acquired during large-scale pretraining.

\subsubsection{Training and Deployment Cost Analysis}
To evaluate the training and deployment costs, we report the FLOPs , the total number of parameters, and the number of trainable parameters for four models: LLM4AMC, LSTM, GRU, and RNN. The results are summarized in Tab. ~\ref{Tab.TDCA}.

\begin{table}[htbp]
\centering
\scriptsize
\caption{Training and deployment cost analysis}
\label{Tab.TDCA}
\begin{tabular}{ccccc}
\toprule
\textbf{Metric} & \textbf{LLM4AMC} & \textbf{RNN} & \textbf{LSTM} & \textbf{GRU} \\
\midrule
Total Parameters(M) & 226.116  & 0.039 & 0.137 & 0.104 \\
Trainable Parameters (M) & 0.513 & 0.039 & 0.137 & 0.104 \\
FLOPS (G) & 3.840 & 0.041 & 0.142 & 0.108 \\
\bottomrule
\end{tabular}
\end{table}

As shown in Tab. ~\ref{Tab.TDCA}, LLM4AMC exhibits the highest total number of parameters and FLOPs, reflecting its large-scale backbone and relatively high deployment cost. However, by freezing most pretrained weights, only 0.513M parameters need to be fine-tuned, which significantly reduces the training cost. In contrast, the recurrent baselines are lightweight models with much smaller parameter sizes and FLOPs, resulting in low deployment overhead. Nevertheless, these models lack the representation capacity of LLM4AMC, leading to weaker overall performance in complex scenarios. This advantage becomes particularly evident in ultra-high-speed scenarios.
In summary, LLM4AMC is efficient during training but costly in deployment, while RNN-based models are lightweight in deployment but limited in performance. In the future, we will explore more lightweight deployment strategies to better balance performance and efficiency.

\subsection{Discussion}

It is important to emphasize that although the experiments in this paper are conducted in a downlink AMC scenario with the model deployed on the UE side, the proposed method is equally applicable to uplink AMC, where the model can be deployed at the base station, thus eliminating the reliance on UE computational power. Therefore, the proposed approach is not necessarily restricted to UE-side execution. Moreover, even in the case of UE-side deployment, our method can leverage common lightweight techniques for large language models, such as model quantization, pruning, and knowledge distillation, to effectively reduce storage and computational overhead, thereby ensuring the feasibility of deployment on the user side. In addition, we note that the timing requirement for AMC decisions is not strictly sub-millisecond. According to 3GPP specifications\textsuperscript{\cite{ref41}}, CQI reporting periodicities are typically configured in multiples of slots or subframes, corresponding to several milliseconds or longer. This indicates that inference times of approximately 1 to 2 ms, such as the 1.54 ms observed for Qwen2.5-0.5B, fall well within the practical timing budget for AMC operations.

Furthermore, we highlight the environmental implications of deploying LLMs at scale in cellular networks. Integrating LLMs into wireless systems can significantly enhance communication performance, thereby reducing pilot and signaling overheads and ultimately lowering the overall energy consumption of the network. Although LLMs are inherently energy-intensive, this drawback can be alleviated through techniques such as model lightweighting. Moreover, recent studies\textsuperscript{\cite{ref25}} have demonstrated the potential of LLMs in multi-task scenarios, further reducing the number of models required. Collectively, these advances contribute to the realization of green communications.

\section{Conclusion}
To address the CQI aging issue in AMC systems, we proposed LLM4AMC, a channel quality prediction and AMC optimization method based on fine-tuned LLMs. Without modifying the existing physical layer architecture, our approach solely incorporated a channel quality prediction module at UE, enabling more accurate CQI feedback that reflects real-time channel conditions. The proposed methodology adopted a four-layer neural architecture comprising sequential processing stages. This hierarchical structure ensured efficient feature extraction while maintaining computational tractability through parameter-efficient transfer learning. Through systematic simulation validation, our proposed method demonstrated significant performance advantages over NP, RNN, LSTM, and GRU baseline models. Particularly in high-mobility scenarios (100 km/h), the approach achieved a 61.96\% reduction in link BLER and a 38.06\% throughput improvement compared to the NP model. Furthermore, comprehensive robustness evaluations, including noise immunity tests, few-shot learning experiments, and cross-scenario generalization trials, whichi confirmed the method's practical deployment feasibility. The results indicated that the solution not only excelled under ideal simulation conditions but also exhibited strong potential for real-world system implementation.

\section{Conflict of interest statement}
None declared.





\section*{About the Authors}\footnotesize\vskip 10mm

\parpic{\includegraphics[width=22mm,height=30mm]{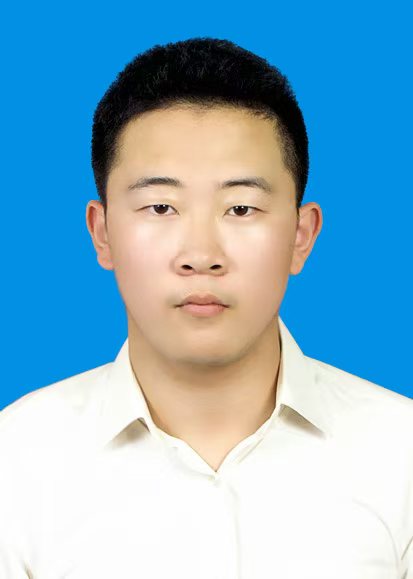}}%
\noindent{\bf Xinyu Pan} received the B.E. degree from Tongji University, Shanghai, China, in 2025. He is currently pursuing the M.S. degree with the School of Electronics, Peking University, Beijing, China. His current research interests include large language model and integrated sensing and communication.
\vskip 4.59mm

\parpic{\includegraphics[width=22mm,height=30mm]{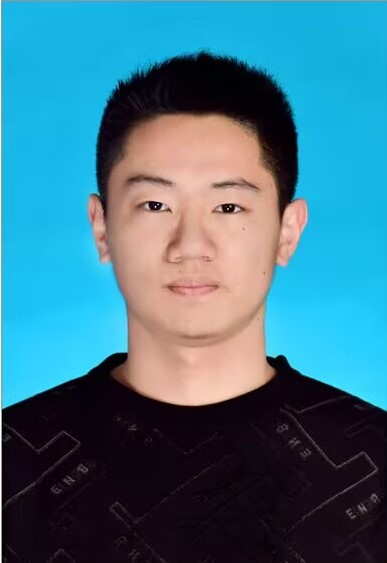}}%
\noindent{\bf Boxun Liu} received the B.E. degree from University of Electronic Science and Technology of China, Chengdu, China, in 2023. He is currently pursuing the Ph.D. degree with the School of Electronics, Peking University, Beijing, China. His current research interests include AI-empowered wireless system design and integrated sensing and communication.
\vskip 4.59mm

\parpic{\includegraphics[width=22mm,height=30mm]{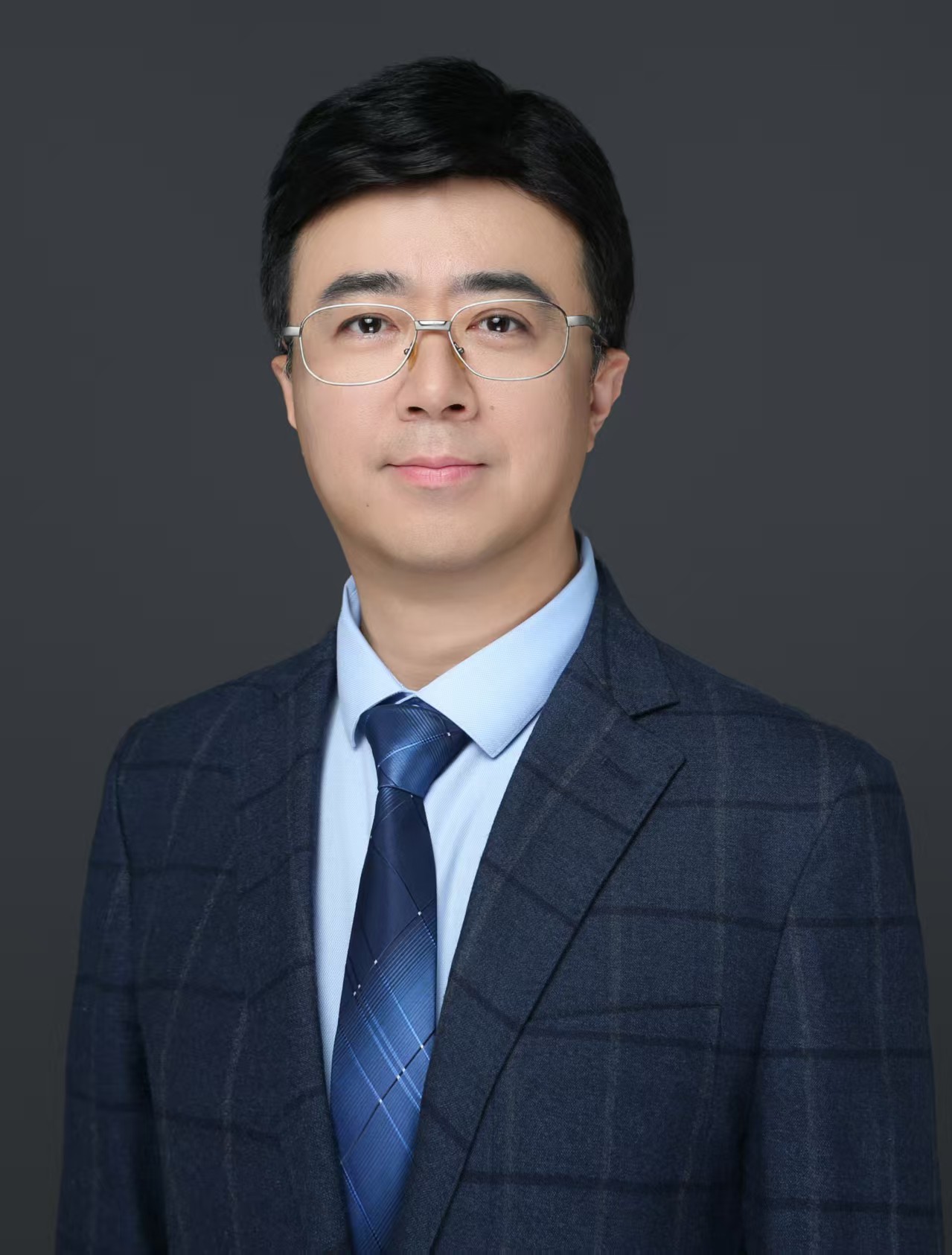}}%
\noindent{\bf Xiang Cheng} [corresponding author] received the joint
Ph.D. degree from Heriot-Watt University and The
University of Edinburgh, Edinburgh, U.K., in 2009.
He is currently a Boya Distinguished Professor with
Peking University. His research focuses on the in-depth integration of communication networks and
artificial intelligence, including intelligent communication networks and connected intelligence, the subject on which he has published more than 280 journals and conference papers, 11 books, and holds 32
patents. He was a recipient of the IEEE Asia–Pacific
Outstanding Young Researcher Award in 2015 and the Xplorer Prize in 2023.
He was a co-recipient of the 2016 IEEE JOURNAL ON SELECTED AREAS
IN COMMUNICATIONS Best Paper Award: Leonard G. Abraham Prize and
the 2021 IET Communications Best Paper Award: Premium Award. He has
also received the Best Paper Awards at IEEE ITST’12, ICCC’13, ITSC’14,
ICC’16, ICNC’17, GLOBECOM’18, ICCS’18, and ICC’19. He has been a
Highly Cited Chinese Researcher since 2020. In 2021 and 2023, he was
selected into two world scientist lists, including the World’s Top 2\% Scientists
released by Stanford University and top computer science scientists released
by Guide2Research. He has served as the symposium lead chair, the co-chair,
and a member of the technical program committee for several international
conferences. He led the establishment of four Chinese standards (including
industry standards and group standards) and participated in the formulation
of ten 3GPP international standards and two Chinese industry standards.
He is currently a Subject Editor of IET Communications; an Associate
Editor of IEEE TRANSACTIONS ON WIRELESS COMMUNICATIONS,
IEEE TRANSACTIONS ON INTELLIGENT TRANSPORTATION SYSTEMS, IEEE WIRELESS COMMUNICATIONS LETTERS, and Journal of
Communications and Information Networks. He was a Distinguished Lecturer
of the IEEE Vehicular Technology Society.

\vskip 4.59mm

\parpic{\includegraphics[width=22mm,height=30mm]{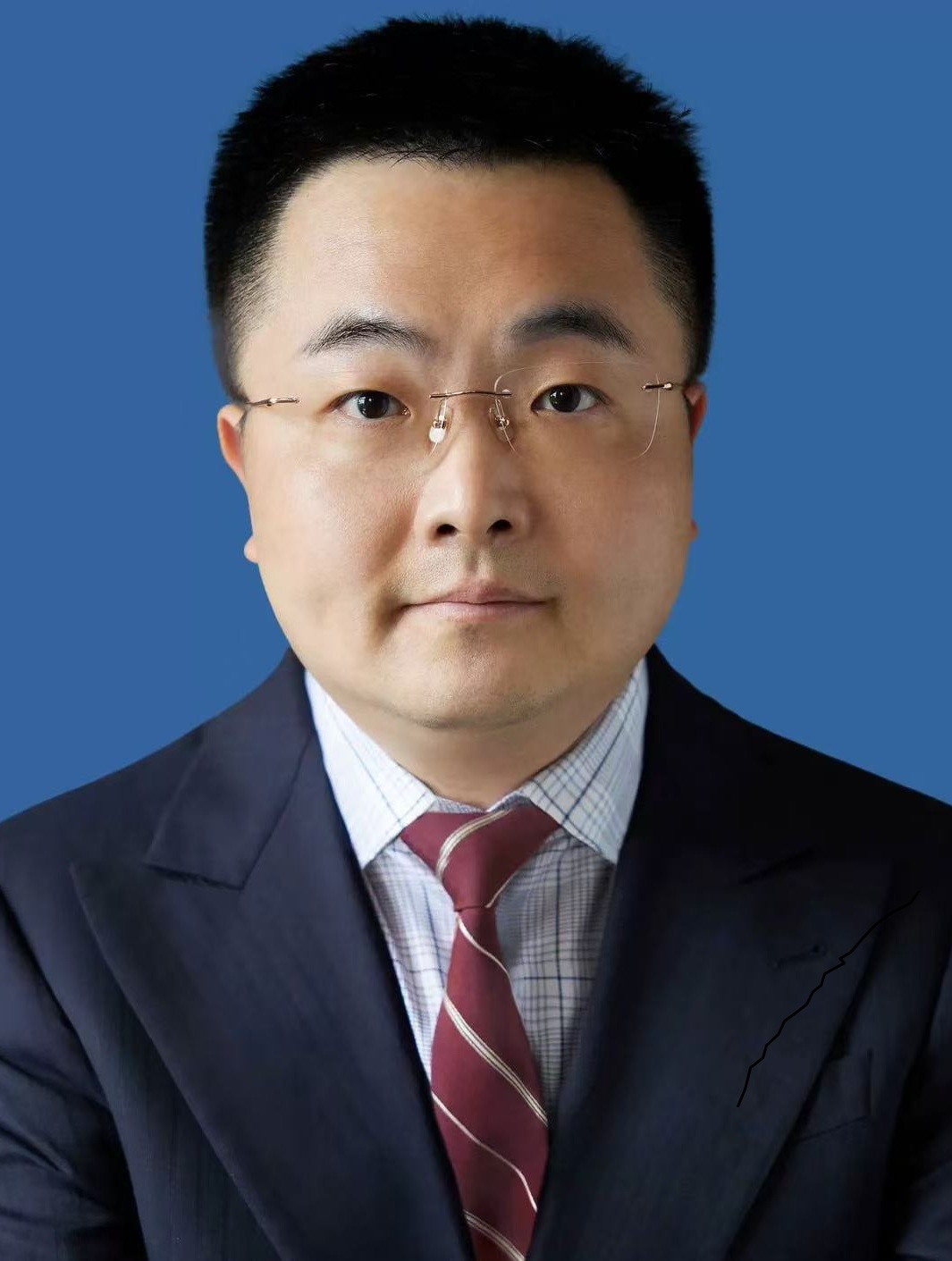}}%
\noindent{\bf Chen Chen}  is currently a Research Professor with Peking University. Since 2010, he has authored or co-authored 5 books and over 120 journal and conference papers, including one “ESI hot paper” (top 0.1\%) and four “ESI highly cited papers” (top 1\%). His research interests include wireless communications and networking, signal processing, and vehicular and satellite communication systems. Dr. Chen was a recipient of two Outstanding Paper Awards from the Chinese Government of Beijing in 2013 and 2018, respectively, and three Best Paper Awards of IEEE conferences (ICNC’17, ICCS’18, and Globecom’18). He has served as the symposium Co-Chair, Session Chair, and a member of the Technical Program Committee for several international conferences. He is currently an Associate Editor of the IET Communications. He has been the principal investigator of over twenty funded research projects and has led the development of a key subsystem in a 12th-13th Five Year Plan Program of China, which has won a First Prize of State Science and Technology Award.
\end{document}